# Some Experimental Results of Relieving Discomfort in Virtual Reality by Disturbing Feedback Loop in Human Brain

Wei Qionghua[1]  Wang Hui[2]  Wei Qiang[3]

**Abstract**
Recently, great progress has been made in virtual reality(VR) research and application. However, virtual reality faces a big problem since its appearance, i.e. discomfort (nausea, stomach awareness, etc). Discomfort can be relieved by increasing hardware (sensor, cpu and display) speed. But this will increase cost. This paper gives another low cost solution. The phenomenon of cybersickness is explained with the control theory: discomfort arises if feedback scene differs from expectation, so it can be relieved by disturbing feedback loop in human brain. A hardware platform is build to test this explanation. The VR display on a Samsung S6 is blurred while head movement is detected. The effect is evaluated by comparing responses to the Simulated Sickness Questionnaire (SSQ) between a control and experimental condition. Experimental results show that the new method can ease discomfort remarkably with little extra cost. As a result, VR may be used more widely in teaching (like foreign language, medicine). It's also reasonable to expect likewise merits in other VR applications.

## 1 Introduction

Virtual reality has made great progress recently. Many kinds of implementations appeared. Among them, Head Mounted Display (HMD) is an important device. Users can get satisfactory 3D immersion by using HMD. However, many users have reported discomfort due to the prolonged use of HMD. This discomfort may lead to stronger effects, such as nausea, eyestrain, headache, and vertigo(Kennedy et al. 1993). It is crucial to ensure that the VR application does not drop frames or produces delays. It's possible to reduce latency by increasing sensor/display speed, but this lead to higher cost. Consequently, many scholars try to solve the problem by other means.

For example, Kijima and Ojika try to reduce tracking latency by sampling the orientation and position as late in the rendering process as possible (Kijima and Ojika 2002). This allows the rendering of visible image to be based on very up-to date tracking information, but has limited quality as only basic adjustments may be made so late in the rendering process. David J. Zielinski et al. try to mitigate the effects of low frame rates (Zielinski et. al. 2015). In their technique, the low frame rate simulation images are displayed with low persistence by blanking out the display during the extra time such image would be displayed. Although helpful for low frame rate, adding black frames may lead to eyestrain. Xingyao Yu et al. try to reduce simulator sickness caused by the low refresh rate of display in smartphone-based VR system by adding static symbol(cross or Minion logo) on the screen of the smartphone (Yu et al. 2016). But adding symbol introduce extra disturbance. This may be undesirable for teaching. Moreover, the discomfort in VR environment may be caused by simulated motion (Bruck and Watters 2009).

Wang Hui
Email：mwanghui@163.com
[1] Henan University of Chinese Medicine, Zhengzhou, China
[2] Zhengzhou Technical College, Zhengzhou, China
[3] Fuzhou University, Fuzhou, China

The collective set of symptoms associated with the perception of motion when no physical motion exists is known as cybersickness (Kim et al. 2005) or Visually Induced Motion Sickness (Howarth and Hodder 2008). This kind of cybersickness cannot be relieved by previous approaches.

This paper tries to explain discomfort caused by VR environment with control theory, i.e. discomfort will happen if feedback (scene perceived by eye) differs from expectation when head rotates. Consequently, blurring display (disturbing feedback loop) may relieve this kind of discomfort (Wei et al. 2017). In fact, this explanation can also explain some interesting phenomenon observed, i.e. researchers have found that stroboscopic glasses (at the 8Hz level) are effective in reducing nausea (Reschke, Somers and Ford 2006). It's also interesting to point out that Yu et al. think blurring display can cause discomfort more quickly (Yu et al. 2016), which is contrary to our results. The reason lies that they did not know the relation between discomfort and feedback error, so they didn't blur image on correct time.

In our research, a set of hardware is designed to test the new theory(explanation). Experiments are carried out to evaluate the effect of the new strategy. A standard questionnaire, the Simulator Sickness Questionnaire (SSQ) is used to quantify simulator sickness symptoms. SSQ is used since other studies have attempted to relate characteristic changes in the physiology of cybersickness with user responses on the SSQ, finding strong correlations between some self-report and physiological measures, so SSQ alone may be sufficient to determine whether a user is suffering cybersickness (Kim et al. 2005).

## 2 The view of control loop

Many factors can cause discomfort in VR environment. Among them, latency and simulation motion are two main causes that are difficult to deal with. But the discomfort caused by these two factors can be explained in a view of control theory. Fig.1 is the structure of general control system, which includes sensor, main processor, actuator, object to be controlled. The principle of the system is very simple: the sensor measures a quantity of the object to be controlled, the processor compare the quantity with a desired value. The processor then calculates an order and sent it to the actuator. This way, the quantity of the object is kept on a desired value.

The perception and action of human being also conform to this pattern. For a human being, the brain is the processor, the muscle that can move the head is actuator, the head is the object to be controlled, and the eyes are sensor. Generally speaking, if the brain wants to change view, it will give order to muscles to rotate the head (or move the eye). At last, the scenes eye perceived should be the same as desired. Moreover, the scene perceived should coincide with the scene expected during the process of head rotating. Discomfort may occur if those two scenes differ from each other.

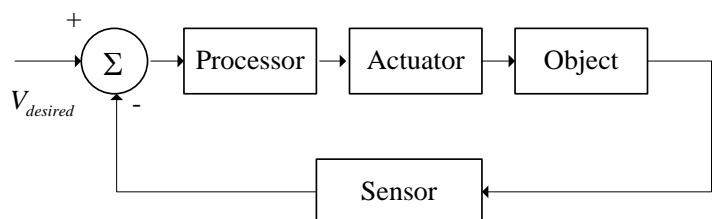

Fig.1 The structure of general control system

Since human beings need to measure distance of object and use tools, the movement of head and eye can be controlled precisely. And the speed of this control-feedback loop is very fast. The brain may expect a change of view, and it can get the corresponding view after muscle action. It's difficult to fulfill this fast/precise feedback loop under current technique level. And the fulfillment inevitably need complex implementation and high cost.

However, there's another possible solution to the discomfort problem. By disturbing feedback loop in human brain, discomfort and cybersickness can be relieved in VR environment. This is practical especially in VR teaching and training. For example, teacher talks with students in a fixed site when VR is used as background for language teaching. There's no need for fast movement or quick action/response then. At that time, a technique in film industry can be used, i.e. The viewpoint can jump from one point/direction to another (even though it remains first person point of view). The continuous transition can be removed by VR devices. In this way, the relation between feedback error and cybersickness can be broken and discomfort can be avoided.

However, there's a shortcoming to this technique. When human beings are watching films, they are in a totally passive mode. The viewpoints are determined by director only. While in VR environment, a user wants to determine viewpoint/direction himself. Disturbing feedback loop will interfere with his will. So the feedback loop should be kept partially. As a result, the scene in user's eyes should be blurred only when his head/eyes moving. The scene should change with his movement, so that the user can stop the movement when the scene agrees with his expectation.

Note that, quickness is also needed in this method. The scene displayed should be blurred as quickly as possible when human head movement is detected. Latency will inevitably affect the performance. Although decreasing latency is a relatively simple task for blurring, it's still a challenge for non real time operation system (like android used in smartphone). So implementing the method totally by software in VR devices is not a good choice. It may be more convenient (and cheap) to implement the method by extra hardwares. This is also the method used in our research.

## 3 Hardware and experiment design

3.1 Hardware

Fig.2 is the structure of the hardware platform used in our research. The core of the platform is a STM32F4 processor. The motion detector is a MPU6050 gyroscope, which is spliced on a side of a VR glasses. It is used to detect motion of user's head. The STM32F4 processor communicates with MPU6050 by means of IIC protocol at the speed of 200 kHz clock frequency. The processor requests and gets angular speed data from gyroscope every 1ms. The processor will generate pulse and send pulse to four MOSFET (H-bridge) to generate 1kHz 80Vp-p square wave if all angular speed data in x-y-z direction is smaller than $6.1°/s$. The output of the H-bridge is connected to liquid crystal sheet placed between VR glass lens and a smartphone (a Samsung S6). As a result, the liquid crystal becomes transparent. On the contrary, if angular speed in any direction is bigger than $6.1°/s$, the processor stops generating square wave, the liquid crystal will change to blurred mode at that time(transparency less than 10%).

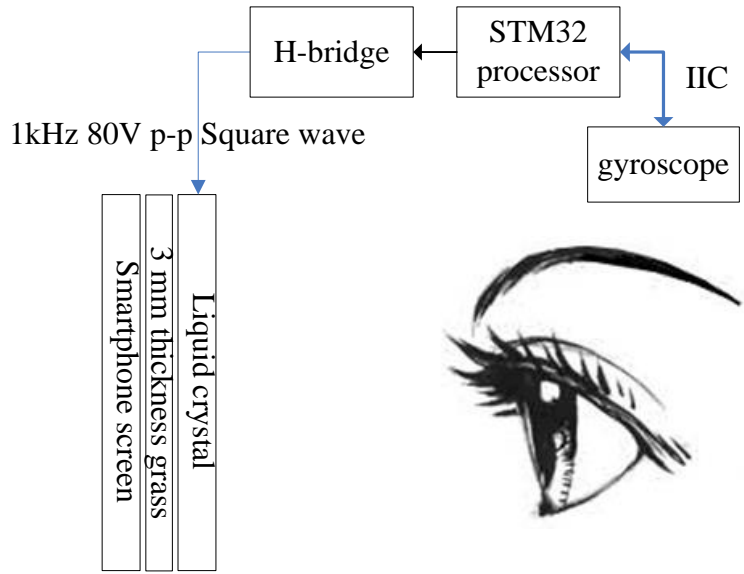

Fig.2 The structure of the hardware platform used (side view)

Since liquid crystal is used to disturb feedback loop when head is moving/rotating, the display as a whole can be blurred when head motion is detected. So a sheet that is larger than smartphone is used. This kind of sheet is used to blur glasses between rooms and outside spaces, so there's no pixels matrix in it. There is only one lead plate on each side of the sheet. Obviously, this kind of liquid crystal sheet is very cheap, and easy to control. However, the display is not vague enough when liquid crystal is not driven by square wave. So a glass whose thickness is 3mm is added between liquid crystal and smartphone screen. In this way, a more vague display can be gotten. Fig. 3 shows the corresponding results.

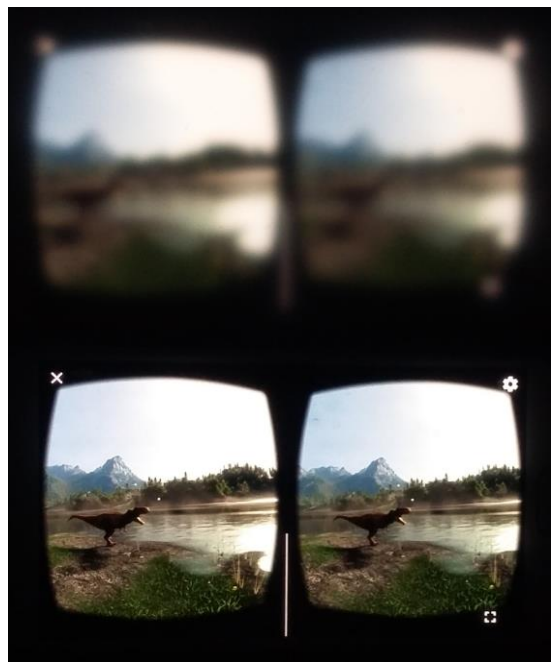

Fig.3 The blurred and clear smartphone display

The display devices used is a Samsung S6 smartphone. The VR glasses is a simple one like Google cardboard. But the focus of two lens and pupillary distance can be adjusted separately. There's no circuit/sensor in the simple VR glasses. Obviously, this is a much cheaper platform compared with some advanced VR HMD (like Oculus Rift).

In this platform, by adding some simple sensor/processor circuit outside smartphone based VR, the feedback loop can be obstructed in less than 2ms after human head motion. This way, nothing (hardware/software) in the smartphone needs to be changed, and discomfort encountered in VR environment may be relieved at much lower cost.

3.2 Software and Experiment Process

The software used during experiment is a VR teaching software designed by Google: Expeditions. This software is different from VR game software. A user can only turn his/her head to possible directions. The user cannot change position (move/fly/walk etc) in the software. This software is selected since it's suitable to evaluate the effect of our method in VR teaching.

Before the experiment, another smartphone (running as guide) download several VR sites in the expedition software. The sites include Taj Mahal/the Thames/Hawaii etc. At the beginning of the experiment, the participants are informed of possible discomfort and danger. They are told that they can discontinue the experiment at any time. They are also told that the display will be blurred while head is moving. After that, the participants are seated in a chair, wearing the VR glasses with SAMSUNG S7/MPU6050/liquid crystal on it. The guide then loads one site after another. In each site, the guide asks the participants to look at several objects in turn. The guide may explain something to the participants, but all commentaries end in 20 seconds. So the participants turn their heads continuously during the experiments. The process lasts for 20 minutes, or ends when a participant discontinues the experiment due to discomfort. After that, the participant is asked to answer SSQ in table 1.

All participants are asked to take part in another experiment (control condition) the next day. The experiment process is the same as in the previous day, but there's no liquid crystal in VR glasses. Besides, randomly selected different sites are used during the experiment, so that participants are not weary. The experiment also lasts for 20 minutes, or ends when a participant discontinue the experiment due to discomfort. After that, the participant is asked to answer SSQ in table 1.

Table 1 SSQ used in experiment

| Symtoms | 0 | 1 | 2 | 3 |
|---|---|---|---|---|
| GeneralDiscomfort | - | - | - | - |
| Fatigue | - | - | - | - |
| Headache | - | - | - | - |
| EyeStrain | - | - | - | - |
| DifficultFocusing | - | - | - | - |
| Nausea | - | - | - | - |
| DifficultyConcentrating | - | - | - | - |
| StomachAwareness | - | - | - | - |
| BlurredVision | - | - | - | - |
| IncreasedSalivation | - | - | - | - |
| DizzyEyesOpen | - | - | - | - |
| DizzyEyesClosed | - | - | - | - |
| Vertigo | - | - | - | - |
| FullnessOfHead | - | - | - | - |

# 4 Experimental results and analysis

## 4.1 Experimental results

The results of the responses to the SSQ in both the experimental and control conditions were analyzed using the parametric matched samples t-test. The results for the Matched Samples t-test can be viewed in Table 2. Significant decreases between the experimental and the control conditions were revealed for general discomfort, eyestrain, nausea, stomach awareness and dizzy (eyes open).

## 4.2 Analysis on some phenomena

a) According to section III.B, the principle of the platform was explained to participants before experiment. So they knew the relationship between blurred vision and head motion. They also knew the vision would become clear again after head stop moving. However, this process was forgotten for one participant. As a result, the participant felt very discomfort (nausea) in 3 minutes. The experiment had to be stopped. After the participant was told about the relationship an hour later, the participant did not feel discomfort during the latter experiment and the experiment ended after 20 minutes. This example may show that the error in feedback loop is a direct cause of discomfort, and the consciously interfering of human brain can play an important role in this loop.

Table 2 MATCHED SAMPLES T-TEST

| Symptom Pairs(blurred-clear) | Mean | Std. Deviation | Std. Error Mean | 95% Confidence Interval of the Difference Upper | 95% Confidence Interval of the Difference Lower | t | df | Sig. (2-tailed) |
|---|---|---|---|---|---|---|---|---|
| GeneralDiscomfortB1 - GeneralDiscomfort1 | -0.65 | 0.74516 | 0.16662 | -0.99875 | -0.30125 | -3.901 | 19 | 0.001 |
| FatigueB2 - Fatigue2 | -0.05 | 0.22361 | 0.05 | -0.15465 | 0.05465 | -1 | 19 | 0.33 |
| HeadacheB3- Headache3 | 0 | 0.32444 | 0.07255 | -0.15184 | 0.15184 | 0 | 19 | 1 |
| EyeStrainB4 - EyeStrain4 | -0.4 | 0.59824 | 0.13377 | -0.67999 | -0.12001 | -2.99 | 19 | 0.008 |
| DifficultFocusingB5 - DifficultFocusing5 | -0.1 | 0.30779 | 0.06882 | -0.24405 | 0.04405 | -1.453 | 19 | 0.163 |
| NauseaB7 - Nausea7 | -0.55 | 0.60481 | 0.13524 | -0.83306 | -0.26694 | -4.067 | 19 | 0.001 |
| DifficultyConcentratingB8 - DifficultyConcentrating8 | -0.2 | 0.41039 | 0.09177 | -0.39207 | -0.00793 | -2.179 | 19 | 0.042 |
| StomachAwarenessB9 - StomachAwareness9 | -0.3 | 0.57124 | 0.12773 | -0.56735 | -0.03265 | -2.349 | 19 | 0.03 |
| BlurredVisionB10 - BlurredVision10 | -0.15 | 0.48936 | 0.10942 | -0.37903 | 0.07903 | -1.371 | 19 | 0.186 |
| IncreasedSalivationB11 - IncreasedSalivation11 | -0.1 | 0.30779 | 0.06882 | -0.24405 | 0.04405 | -1.453 | 19 | 0.163 |
| DizzyEyesOpenB12 - DizzyEyesOpen12 | | -0.55 | 0.68633 | 0.15347 | -0.87121 | -0.22879 | -3.584 | 19 | 0.002 |
| DizzyEyesClosedB13 - DizzyEyesClosed13 | | -0.2 | 0.52315 | 0.11698 | -0.44484 | 0.04484 | -1.71 | 19 | 0.104 |
| VertigoB14 - Vertigo14 | | -0.05 | 0.22361 | 0.05 | -0.15465 | 0.05465 | -1 | 19 | 0.33 |
| FullnessOfHeadB16 - FullnessOfHead16 | | -0.15 | 0.36635 | 0.08192 | -0.32146 | 0.02146 | -1.831 | 19 | 0.083 |

b) Three participants discontinued the experiment in control condition due to unbearable discomfort. On the contrary, all participants finished the 20-minute experiments when blur effect is used. Although most of them did not feel discomfort after the experiment, some did report nausea or eyestrain. We think these discomfort may be caused by other factors like pupillary distance or VR glass profile mismatch. For example, the pupillary distance of one participant is much smaller than others. She couldn't compose 3D view in her brain when default pupillary distance was used. The experiment couldn't be carried out until correct pupillary distance was used. Besides, since transparent glasses were added to ensure blur effect, the Gear VR which is compatible with SAMSUNG s6 cannot be used, so there's no precise match between VR glass profile and profile used in smartphone. In fact, it has been reported that many factors can lead to discomfort in VR environment (Thiago et al. 2017). We believe that all kinds of discomfort can be eliminated by using our method and eliminating other factors.

## 5 Conclusion

The discomfort felt in VR environment can be explained with control theory. If view gotten by sensor (eye) differs from expectation, discomfort may be aroused. Disturbing this feedback loop can relieve discomfort. A platform is build upon this principle. Extra hardware including gyroscope, processor and liquid crystal are added to VR glasses and smartphone. The STM32 processor can stop sending square wave to liquid crystal in 2ms after detecting head motion. As a result, the display on smartphone is blurred and the feedback is disturbed. The platform and Google VR software expedition are used to test the principle proposed. Twenty participants take part in the experiment in experimental and control conditions. SSQ is used to quantify the degree of discomfort and parametric matched samples t-test is used to evaluate the effect of the new principle. The results show that the new principle can relieve discomfort effectively.

The implementation method used in this paper can be used on low-cost smartphone since it is very cheap. This may increase the chance of VR application in teaching greatly. Note, however, that the principle can also be used in many other VR applications. For example, users need to walk or fly in VR games. In those games, the view may be changed but there's no corresponding physical movement. It's less likely that the discomfort caused may be relieved by using better hardware or decreasing latency. But the principle proposed in this paper may be useful at that time. This is also the area being explored now.

## Reference

Bruck Susan and Watters Paul A. (2009) Estimating Cybersickness of Simulated Motion Using the Simulator Sickness Questionnaire (SSQ): A Controlled Study , 2009 Sixth International Conference on Computer Graphics. Imaging and Visualization, pp 486 - 488

Howarth P. A. and Hodder S.G. (2008)